\documentclass{aa}

\usepackage{graphicx}
\usepackage{amssymb}
\usepackage{latexsym}
\usepackage{graphics}
\usepackage[varg]{txfonts}

\makeatletter
\renewcommand\@biblabel[1]{}
\makeatother

\def\ha{H$\alpha$}
\def\hb{H$\beta$}

\def\oii{[OII]$\lambda$3727}
\def\niib{[NII]$\lambda$6584}

\def\oiiib{[OIII]$\lambda$5007}

\begin{document}


\title{Fundamental metallicity relation in CALIFA, SDSS-IV MaNGA, and high-z galaxies}
        
\author{G. Cresci\inst{1}, F. Mannucci\inst{1} and M. Curti\inst{2,3}
}

\institute{INAF - Osservatorio Astrofisco di Arcetri, largo E. Fermi 5, 50127 Firenze, Italy \\ \email{giovanni.cresci@inaf.it}
\and
Cavendish Laboratory, University of Cambridge, 19 J. J. Thomson Ave., Cambridge CB3 0HE, UK 
\and
Kavli Institute for Cosmology, University of Cambridge, Madingley Road, Cambridge CB3 0HA, UK  
}

\date{Received ; accepted }

\abstract{The metallicity of local galaxies is tightly related not only to stellar mass, i.e. the mass-metallicity relation, but also to the star formation rate (SFR) through the so-called fundamental metallicity relation (FMR); more active galaxies show lower metallicities at fixed mass. Interestingly, high-z galaxies up to $z\sim2.5$ follow the same relation defined by SDSS locally. However, different shapes have been proposed for local galaxies, and the existence of a FMR and the role of the SFR has been recently questioned by some authors.
In this paper we first discuss the various parametrizations of this mass-metallicity-SFR relation  that has appeared in the literature to understand the origin of their different shapes. We then reanalysed data from CALIFA and SDSS-IV MaNGA surveys, which were used to suggest no dependency of metallicity on the SFR in local galaxies. Contrary to those claims, we find that those datasets are instead fully consistent with the predictions, showing the expected dependency on the SFR at fixed mass. Finally, we analysed those high-z data whose consistency with the local relation was questioned. While an internal dependency on the SFR among the subsamples is difficult to detect at high-z because of the limited dynamic range sampled in the three parameters and the intrinsic scatter and uncertainties of such small samples, all these datasets are compatible with the relation defined locally by SDSS galaxies. This confirms the lack of evolution of the FMR in these data up to $z\sim2.3$.
}


\authorrunning{Cresci et al.}
\titlerunning{Fundamental metallicity relation in CALIFA, MaNGA and high-z galaxies}

\maketitle

\section{Introduction}

Astronomers often search for correlations between various physical properties of galaxies, as these may indicate important underlying physical relationships between the examined variables, for example the elliptical galaxies fundamental plane (Djorgovski \& Davis \citealp{djorgovski87}), the Tully-Fisher relation (Tully \& Fisher \citealp{tully77}, Cresci et al. \citealp{cresci09}), the black hole mass - bulge mass relation (Magorrian et al. \citealp{magorrian98}), and the main sequence (Brinchmann et al. \citealp{brinchmann04}, Noeske et al. \citealp{noeske07}, Popesso et al. \citealp{popesso18}), etc. \\
Interestingly, it was soon recognized that a tight relation is present between the metal content of galaxies in the gas phase and their magnitude (McClure \& van den Bergh \citealp{mcclure68}; Lequeux et al. \citealp{lequeux79}), or better yet their mass (e.g. Garnett \citealp{garnett02},  Tremonti et al. \citealp{tremonti04}, Lee et al. \citealp{lee06}, Kewley \& Ellison \citealp{kewley08}, Salim et al. \citealp{salim14}, and many others); brighter and more massive galaxies show higher metal enrichment. In particular, Tremonti et al. \cite{tremonti04} used Sloan Digital Sky Survey
 (SDSS) spectroscopy to show that this mass-metallicity (MZ) relation in the local Universe spans at least three orders of magnitude in stellar mass and a factor of ten in metallicity and has a scatter of $\sim0.1$ dex. 

Ellison et al. \cite{ellison08} first suggested that the star formation rate (SFR) of galaxies could have an influence on the metallicity, finding a mild relation between the specific SFR ($sSFR=SFR/M_*$) of galaxies in pairs and their metallicities. Mannucci et al (\citealp{mannucci10}, hereafter M10) showed that all star-forming galaxies form a very tight surface in the 3D space defined by gas-phase metallicity, stellar mass, and SFR, dubbed fundamental metallicity relation (FMR), which reduces the scatter of individual galaxies to only $\sim0.05$ dex, i.e. compatible with the measurement uncertainties on the parameters involved. These authors used the SDSS-DR7 dataset of $\sim 140000$ galaxies with $0.07<z<0.3$, clearly showing a significant dependency of metallicity on SFR at fixed mass, in which more actively star-forming galaxies have lower metallicity. The relation was extended by Mannucci et al. \cite{mannucci11} towards lower masses.
This new relation appears not to evolve in redshift, at least up to $z\sim2.5$, with high-z galaxies following the same relation defined in the local Universe (e.g. Richard et al. \citealp{richard11}, Cresci et al. \citealp{cresci12},  Belli et al. \citealp{belli13}, Maier et al. \citealp{maier14}, Kacprzak, et al. \citealp{kacprzak16}). Therefore, in this framework the observed redshift evolution of the MZ relation, in which high-z galaxies show lower metal content (see e.g. Erb et al. \citealp{erb06}, Maiolino et al. \citealp{maiolino08}, Mannucci et al. \citealp{mannucci09}), would be due to their higher SFR (Schreiber et al. \citealp{schreiber15}). However, some evolution is observed at $z>3$, where lower metallicities than those predicted by the FMR are observed (Troncoso et al. \citealp{troncoso14}).

These results were later confirmed by several other groups and galaxy samples, both locally (Lara-L\'opez et al. \citealp{copioni10}, Yates et al. \citealp{yates12}, Mannucci et al. \citealp{mannucci11}, Hunt et al. \citealp{hunt12}, Andrews \& Martini \citealp{andrews13}, Nakajima \& Ouchi \citealp{nakajima14}, Salim et al. \citealp{salim14}, Grasshorn Gebhardt et al. \citealp{grasshorn16}, Cresci et al. \citealp{cresci17}, Jimmy et al. \citealp{jimmy15}, Wu et al. \citealp{wu16}, Hirschauer et al. \citealp{hirschauer18}) and at high redshift (Richard et al. \citealp{richard11}, Nakajima et al. \citealp{nakajima12}, Cresci et al. \citealp{cresci12}, Magrini et al. \citealp{magrini12}, Niino et al. \citealp{niino12}, Christensen et al. \citealp{christ12}, Belli et al. \citealp{belli13}, P{\'e}rez-Montero, et al. \citealp{pm13}, Stott et al. \citealp{stott13}, Cullen et al. \citealp{cullen14}, Yabe et al. \citealp{yabe14},  Maier et al. \citealp{maier14}, Divoy et al. \citealp{divoy14}, Stott et al. \citealp{stott14}, Yabe et al. \citealp{yabe15},   Salim et al. \citealp{salim15}, Kacprzak, et al. \citealp{kacprzak16}, Calabr\'o et al. \citealp{calabro17}, among others), although in some cases an offset from the $z\sim0$ relation was found, suggesting a redshift evolution (Zahid et al. \citealp{zahid14}, Sanders et al. \citealp{sanders18}, Gao et al. \citealp{gao18}, Pharo et al. \citealp{pharo18}).\\

As already suggested in Ellison et al. \cite{ellison08} and M10, the existence of the 3D relation can be explained by the interplay of infall of pristine gas and outflow of enriched material (e.g. Dav{\'e}, et al. \citealp{dave11}, Dayal et al. \citealp{dayal13}, Lilly et al. \citealp{lilly13}, Forbes et al. \citealp{forbes14}, Harwit \& Brisbin \citealp{harwit15}, De Rossi et al. \citealp{derossi15}, \citealp{derossi17}, Torrey et al. \citealp{torrey18}). For this reason, the FMR is expected to be just a proxy of a primary relation between the metallicity, mass, and gas content of galaxies. Such relation has been in fact observed using HI content by Bothwell et al. \cite{bothwell13}, Hughes et al. \cite{hughes13}, Brown et al. \cite{brown18}, and by Bothwell et al. \cite{bothwell16} using molecular gas.

In some cases the correlation of metallicity with both mass and SFR is present, but the overall shape of the correlation changes significantly (see e.g. Lara-L\'opez et al. \citealp{copioni10}, Yates et al. \citealp{yates12}, Hunt et al. \citealp{hunt12}). As discussed in Sect.~\ref{shape}, these effects are due to the different sample selection or to the assumed calibration and methodology to derive SFR and metallicities, while the dependency on SFR of the measured metallicity at fixed mass is very robust and has been confirmed.\\

However, the existence of the FMR and the role of the SFR has been questioned by some authors, 
both locally (S\'anchez et al. \citealp{sanchez13}, de los Reyes et al. \citealp{delos15}, Barrera-Ballesteros et al. \citealp{barrera17}, S\'anchez et al. \citealp{sanchez17}) and at high-z (Steidel et al. \citealp{steidel14}, Wuyts et al. \citealp{wuyts14}, Sanders et al. \citealp{sanders15}). 
In Sect. 2 we discuss the origin of the different shapes of the FMR obtained by different authors. We then discuss the analysis presented in some of the works above for galaxies in the local Universe (in Sect. \ref{sanchez} and \ref{barrera}) and at high-z (Sect.~\ref{highz}) to show that the FMR is actually consistent with those data. Our conclusions follow in Sect. \ref{conclusion}.

\section{Different shapes of the FMR} \label{shape}

After the seminal work of M10, the shape and parametrization of the FMR were investigated by several groups, each applying their own sample selection, as well as metallicity and SFR calibrations. In this Section we briefly review some of these different parametrizations to better understand the underlying differences.
\begin{figure}
        \begin{center}
                \includegraphics[width=0.45\textwidth]{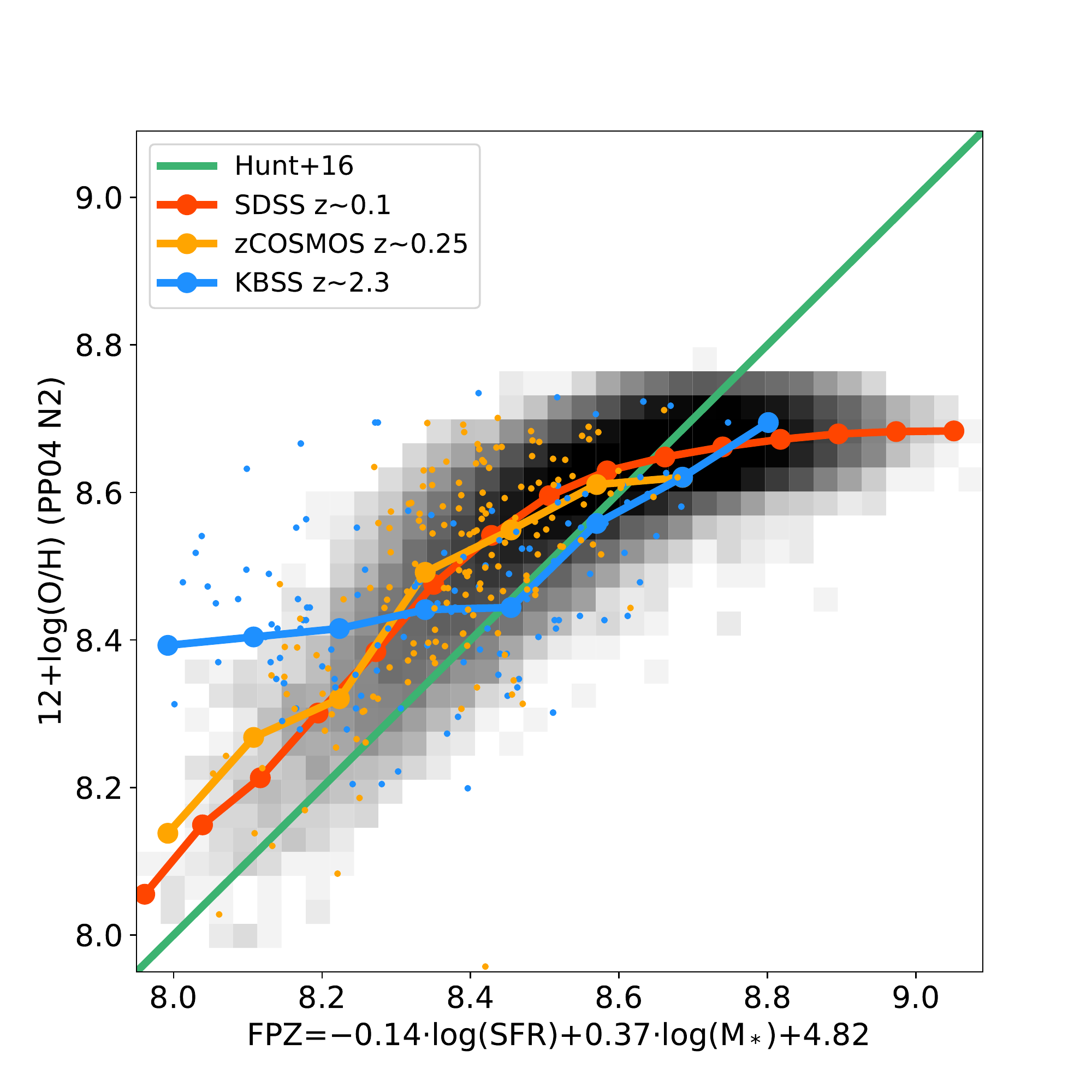}
        \end{center}
        \caption{So-called fundamental plane by Hunt et al. \cite{hunt16}, as derived by PCA of a sample of $\sim1000$ low metallicity starbursts, shown as a green line. The distribution of SDSS galaxies from M10 is overplotted with their median metallicity values in mass bins shown in red. The high-z samples with a significant number of galaxies and $z<3$ and measured N2 index used by Hunt et al. \cite{hunt16} are also reported; the zCOSMOS galaxies at $z\sim0.25$ from Cresci et al. \cite{cresci12} are indicated in orange and the KBSS $z\sim2.3$ galaxies from Steidel et al. \cite{steidel14} in blue. All metallicities have been derived with the same calibration based on N2 (Pettini \& Pagel \citealp{pettini04}). Small dots represent the single galaxy measurements, while the larger circles the medians in mass bins. The general population of local star-forming galaxies, as traced by SDSS, as well as the higher z samples are not well reproduced by the plane.}
        \label{fighunt}
\end{figure}

Lara-L\'opez et al. \cite{copioni10} and Hunt et al. \cite{hunt12} both described as a plane the locus populated by galaxies in the $M_*$-SFR-Z space, instead of a curved 3D surface as in M10. They both used principal component analysis (PCA) or linear regression to derive the best-fitting plane. \\
In particular, Lara-L\'opez et al. \cite{copioni10} used SDSS-DR7 galaxies as M10, but limited their analysis to galaxies with $0.04<z<0.1$ to work with a complete sample in magnitude and redshift. In contrast, the authors of M10 used galaxies with $z>0.07$ and median redshift $<z>=0.106$ to ensure that \oii\ was within the useful spectral range and that $3"$ aperture of the spectroscopic fibre samples a significant fraction of the galaxy ($\sim4$ kpc at $z=0.07$). The Lara-L\'opez et al. \cite{copioni10} upper cut in redshift, however, results in a poorer sampling of the high mass population, and thus most galaxies on the flat part of the MZ at high metallicity are missing. 
Moreover, one of the original aims of the paper is to estimate `stellar masses from SFR and metallicity'. Therefore, contrary to all other studies including M10, these authors bin galaxies as a function metallicity and SFR, deriving the median mass in each bin. They fitted a plane to this distribution of masses, which they called the fundamental plane, which is quantitatively very different from the FMR and with a much larger scatter, $0.32$ dex in mass.
They confirmed the result by M10 that this 3D relation or plane does not significantly evolve with redshift, using comparison samples up to $z\sim3$. This fundamental plane was later revised in Lara-L\'opez et al. \cite{copioni13}, where they refitted a plane to SDSS data now extended to higher redshifts and masses with linear regression and PCA, confirming the correlation between $M_*$, SFR, and Z. They obtained a scatter $\sigma=0.1$ around their best-fitting plane, given by linear regression of $log(M_*)$ as a function of a combination SFR and metallicity. Even though this scatter is larger than the M10 FMR ($\sigma=0.05$) and consistent with the original MZ value, they claimed that the FMR do not provide the best representation of the data, although their fit was actually compared with the work of Yates et al. \cite{yates12} (see below) and not directly with M10.\\
\begin{figure*}
        \begin{center}
                \includegraphics[width=1\textwidth]{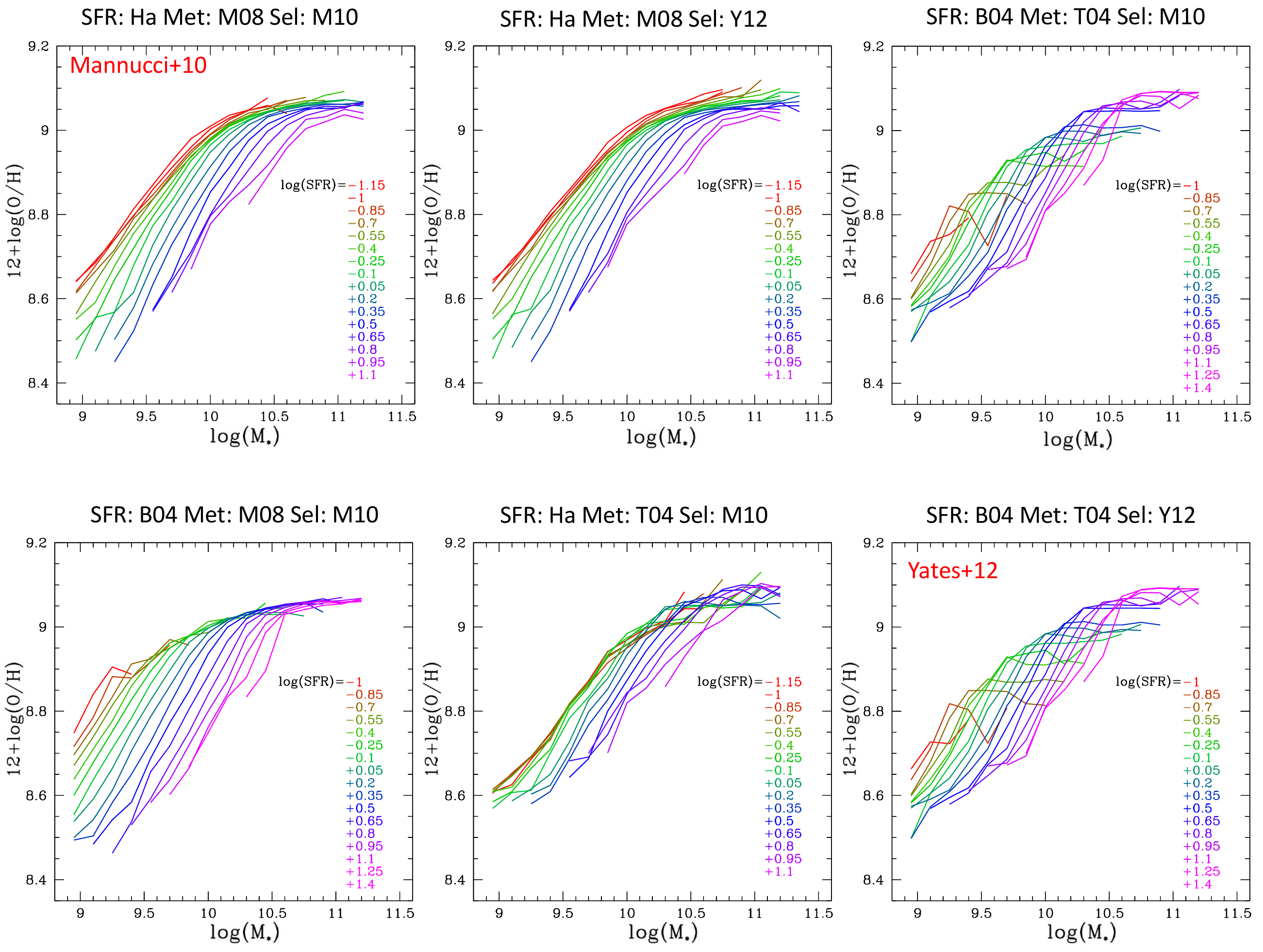}
        \end{center}
        \caption{For SDSS galaxies, MZ relation binned in SFR. The panels show different combinations of selection (either from M10 or Yates et al. \citealp{yates12}, Y12), metallicity (either from Maiolino et al. \citealp{maiolino08}, M08, as in M10 or from Tremonti et al. \citealp{tremonti04}, T04 as in Yates et al. \citealp{yates12}), and SFR (either from H$\alpha$ inside the fibre apertures, as in M10, or from Brinchmann et al. \citealp{brinchmann04}, B04, as in Yates et al. \citealp{yates12}). The upper left diagram corresponds to the M10 work, using the Maiolino et al. \cite{maiolino08} metallicity calibration, and SFR derived from dust-corrected H$\alpha$; the lower right corresponds to Yates et al. \cite{yates12}, which instead uses Brinchmann et al. \cite{brinchmann04} SFR estimates and Tremonti et al. \cite{tremonti04} metallicities. It can be seen how the combination of both a different metallicity calibrator and a SFR calibrator is needed to reproduce the turn off at high mass described in Yates et al.\cite{yates12}, while the selection applied has less impact.} 
        \label{yates}
\end{figure*}

As opposed to M10 and Lara-L\'opez et al. (\citealp{copioni10}, \citealp{copioni13}), Hunt et al. \cite{hunt12} used as reference a heterogeneous sample of $\sim1000$ low metallicity starbursts between $0<z<3.4$ with metallicities derived using different calibrations, of which 803 are drawn from the Izotov et al. \cite{izotov11} luminous compact emission-line galaxies (LCG) sample; these galaxies are characterized by green colours, compact structures, and low metallicities. Even though they were selected to be extreme and rare objects with low metal content, and therefore clear outliers of the MZ relation, the galaxies of this sample are reasonably reproduced by the FMR, although with a scatter of $\sigma=0.27$ dex and an offset of $-0.18$ dex, as discussed in Hunt et al. \cite{hunt12} (see e.g. their Fig.~5). The shift between the LCG sample and the M10 FMR, based on the semi-empirical calibration by Maiolino et al. \cite{maiolino08}, is consistently due to the different metallicity calibration used. In fact, the metallicities in the LCG sample are derived by Izotov et al. \cite{izotov11} using the direct method, obtaining the electron temperature $T_e$ from the ratio of the auroral line [OIII]$\lambda$4363 to \oiiib, that is known to provide lower abundances with respect to the so-called strong-line method calibrated against photoionization models (e.g. Kewley \& Ellison \citealp{kewley08}, Andrews \& Martini \citealp{andrews13}, Curti et al. \citealp{curti17}). The magnitude of the observed shift is of the same order of the difference expected between the two calibrations in the LCG metallicity range ($\sim0.15$ dex, Hunt et al. \citealp{hunt12}, Kewley \& Ellison \citealp{kewley08}). \\
\begin{figure*}
        \begin{center} 
                \includegraphics[width=0.9\textwidth]{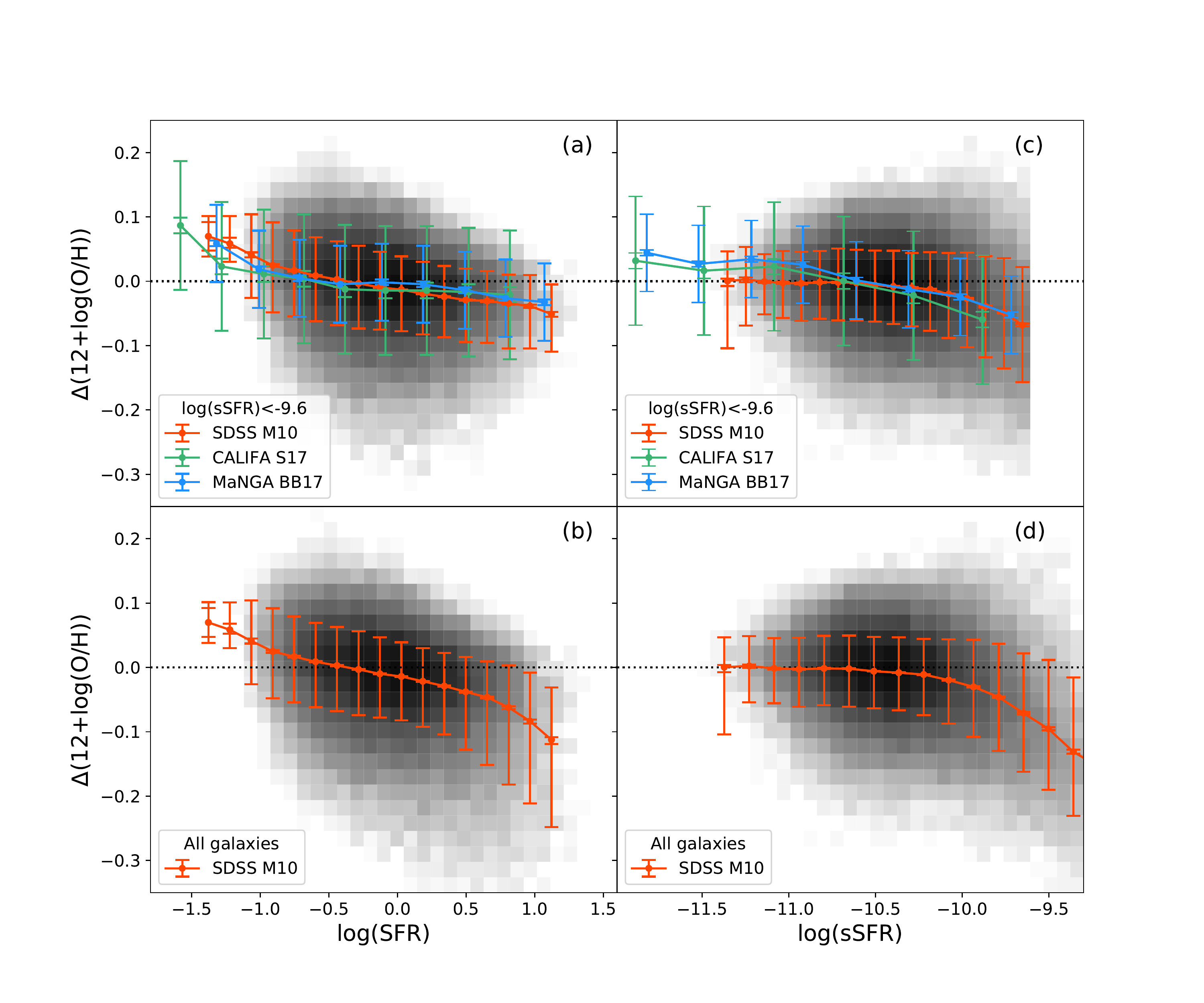}
        \end{center}
        \caption{Difference between the measured metallicity of the SDSS M10 sample artificially cut at $log(sSFR)<-9.6$, to match the S17 and BB17 samples, and the median MZ relation as a function of the SFR (panels a), and as a function of sSFR (panel c). The medians in each SFR and sSFR bin are shown as red points. It can be seen how the expected behaviour for the M10 sample is basically a flat curve, with a mild signal at the lowest $log(SFR)<-1.2$, where higher metallicities $\Delta (log(O/H))<0.1\ dex$ are expected. The same relation as derived for CALIFA galaxies by S17 is overplotted with green points, while the data from MaNGA galaxies derived by BB17 are overplotted as blue points. The error bars shown are the mean standard deviation along the considered bins for MaNGA and CALIFA, as reported by B17 and S17, while they show 1$\sigma$ dispersion in each bin for SDSS. For comparison, we also plot the standard errors on the mean for the three samples. In panels b and d the same plots are reported but for the full SDSS M10 sample, without a sSFR cut. In this case, metallicities lower by $\Delta (log(O/H))<0.1$ dex are found at the highest $log(SFR)>0.8$. All panels clearly demonstrate that most of the dependency of the MZR residuals on SFR is hidden when the sample is not split into mass bins. 
        }
        \label{allmass}
\end{figure*}
Hunt et al. \cite{hunt12} applied a PCA to the galaxies in their sample and derive a plane in the SFR-$M_*$-Z space that best reproduces their distribution, finding a scatter of $\sigma=0.17 $ dex for their starburst sample around this plane. However, the general local galaxy population traced by SDSS is not reproduced by this plane and appears to show an offset of $\sim0.4$ dex (see their Fig.~7). The dispersion obtained by Hunt et al. \cite{hunt12} for SDSS galaxies was similar to the FMR dispersion ($\sigma=0.06$), although in this case the SDSS sample was artificially cut by Hunt et al. \cite{hunt12}  above $log(M_*)>10.5$, to reproduce with a plane the curved galaxy distribution, which shows a bent and a metallicity plateau above that mass.\\
Hunt et al. \cite{hunt16} recomputed the PCA using a different sample of galaxies including a larger fraction of high-z objects, mostly zCOSMOS galaxies at $z\sim0.6$ from Cresci et al. \cite{cresci12} and the $z\sim2.3$ galaxies from Steidel et al. \cite{steidel14}, now rescaled to a common metallicity calibration. Also in this case their simple plane, called  the fundamental plane of metallicity (FPZ) is clearly not able to reproduce the curved shape of the 3D galaxy distribution (see Fig.~\ref{fighunt}). The final scatter considering the medians in mass bins of SDSS galaxies from Fig.~\ref{fighunt} is $\sigma\sim0.15$ dex, i.e. considerably larger than the 0.001 dex scatter of mass bins in the original FMR by M10. Moreover, a shift of $\sim 0.11$ dex is also present between the plane and the high-z samples, which is consistent with the  0.16 dex shift between their `MEGA' and $z>2$ samples already found in their Fig.~7 by Hunt et al. \cite{hunt16}, despite the claim of redshift invariance of the FPZ. \\

Yates et al. \cite{yates12} reanalysed the SDSS-DR7 sample used by M10, but made different choices for sample selection and parameters estimation. In particular, they used the Brinchmann et al. \cite{brinchmann04} fibre-corrected total SFR derived from aperture-corrected emission line fluxes and a Bayesian technique based on population synthesis models from Bruzual \& Charlot \cite{bc03} and HII regions models from Charlot \& Longhetti \cite{charlot01}, instead of the H$\alpha$ based SFR of M10 inside a fibre aperture. Moreover, the metallicities were computed using the Tremonti et al. \cite{tremonti04} procedure based on a Bayesian fit of the main emission lines with a grid of photoionization models, while M10 used the semi-empirical N2=[NII]/H$\alpha$ and R23=([OIII]$\lambda$4958,5007+[OII])/H$\beta$ calibrations from Maiolino et al. \cite{maiolino08}. Finally, the sample selection cut was fixed in Yates et al. \cite{yates12} so that S/N(\ha, \hb, \niib)$>5$ and SNR(\oiiib)$>3$. This is in contrast with M10 who fixed only S/N(\ha)$>25$ to ensure that all the main optical emission lines are generally detected with enough S/N without introducing metallicity biases. Yates et al. \cite{yates12} already noticed that these different choices translate into different shapes for the resulting FMR. They obtained a clear dependency of the MZ on SFR, but the trend is opposite for low and for high stellar mass galaxies; low mass galaxies with high SFR show lower metallicity at fixed mass and high mass galaxies with high SFR instead show higher metallicity. In practice, the different MZ relations for galaxies at fixed SFR are crossing at $log(M_*)\sim10.2$, while above that mass the M10 FMR shows only a very mild dependency on SFR (see Fig.~\ref{yates}, upper left and lower right panels). Yates et al. \cite{yates12} (see also Lara-L\'opez et al. \citealp{copioni13}) ascribed the different shape of their relation to the different method used to derive the metallicity. In Fig.~\ref{yates} we show how the different combinations of selection, SFR, and metallicity determination used in these two works actually affect the shape of the FMR. We confirm that the choice of the metallicity calibration has an influence on the shape of the relation, thereby producing an inversion at high masses (see Fig.~\ref{yates}, lower centre panel). However, the effect is even stronger when the Tremonti et al. \cite{tremonti04} metallicities are coupled with the Brinchmann et al. \cite{brinchmann04} aperture-corrected SFR estimates (Fig.~\ref{yates}, right panels). In this case the inversion appears already at $log(M_*)>9.5$ for the lowest SFR bin. A deep analysis of the advantages and disadvantages of these different choices for parameter estimation is beyond the scope of this paper. We just note that aperture correction for total SFR computation might be very uncertain, especially for massive galaxies hosting large bulges (e.g. Brinchmann et al. \citealp{brinchmann04}, Duarte Puertas et al. \citealp{duarte17}) and that MZ relations built using direct $T_e$ based metallicities generally do not show the inversion reported by Yates et al. \cite{yates12} (see e.g. Andrews \& Martini \citealp{andrews13}, Curti et al. in prep.). \\

Summarizing, despite the differences in the shape of the 3D relation between $M_*$, SFR and gas metallicity, all these works confirm that these three properties are closely connected, and that the scatter of the MZ relation can be reduced by using the SFR as third parameter in the local as well as in the higher-z Universe. Given the 3D shape of the relation, a simple linear regression does not reproduce the observed data. In the following sections, we reanalyse the few datasets that were used to claim that such  a three-parameter relationship does not hold.

\section{Fundamental metallicity relation in CALIFA galaxies} \label{sanchez}

The Calar Alto Legacy Integral Field Area survey (CALIFA; S\'anchez et al. \citealp{sanchez12}) has made use of the PMAS/PPAK spectrometer to obtain  spatially  resolved  spectroscopic  information  of  a  diameter  selected  sample  of $\sim700$  galaxies  in  the  local Universe ($0.005<z<0.03$). S\'anchez et al. (\citealp{sanchez13}, hereafter S13) used $\sim3000$ HII regions in the first 150 galaxies of the survey to study their MZ relation. They assigned to each galaxy the metallicity at its effective radius $R_{eff}$, derived interpolating the radial metallicity gradient from single HII regions measurements. In this way, they obtained a MZ relation with a shape consistent with previous results and a small scatter of $\sigma=0.07\ dex$. However, they claimed that the SFR of each galaxy has no influence on this scatter, contrary to FMR expectations. In fact, when all galaxies are plotted together they cannot see any dependency of metallicity on SFR (their Fig.~4, lower right panel). \\
The same data presented in S13 were reanalysed by Salim et al. \cite{salim14}, who instead plotted in their Fig. 10 the sSFR normalized to the MS $\Delta log(sSFR)$ versus metallicity in mass bins, finding a convincing dependency on SFR in all mass bins except the lowest bin, in which too few points were available and the CALIFA survey is incomplete. Salim et al. \cite{salim14} correctly argued that the lack of dependency initially found by S13 was because the sample was not split by stellar mass, and because of its deficiency of extreme star formers and large fraction of low sSFR targets, which makes the dependency on star formation very mild once the entire sample is plotted together. \\
\begin{figure}
        \begin{center} 
                \includegraphics[width=0.5\textwidth]{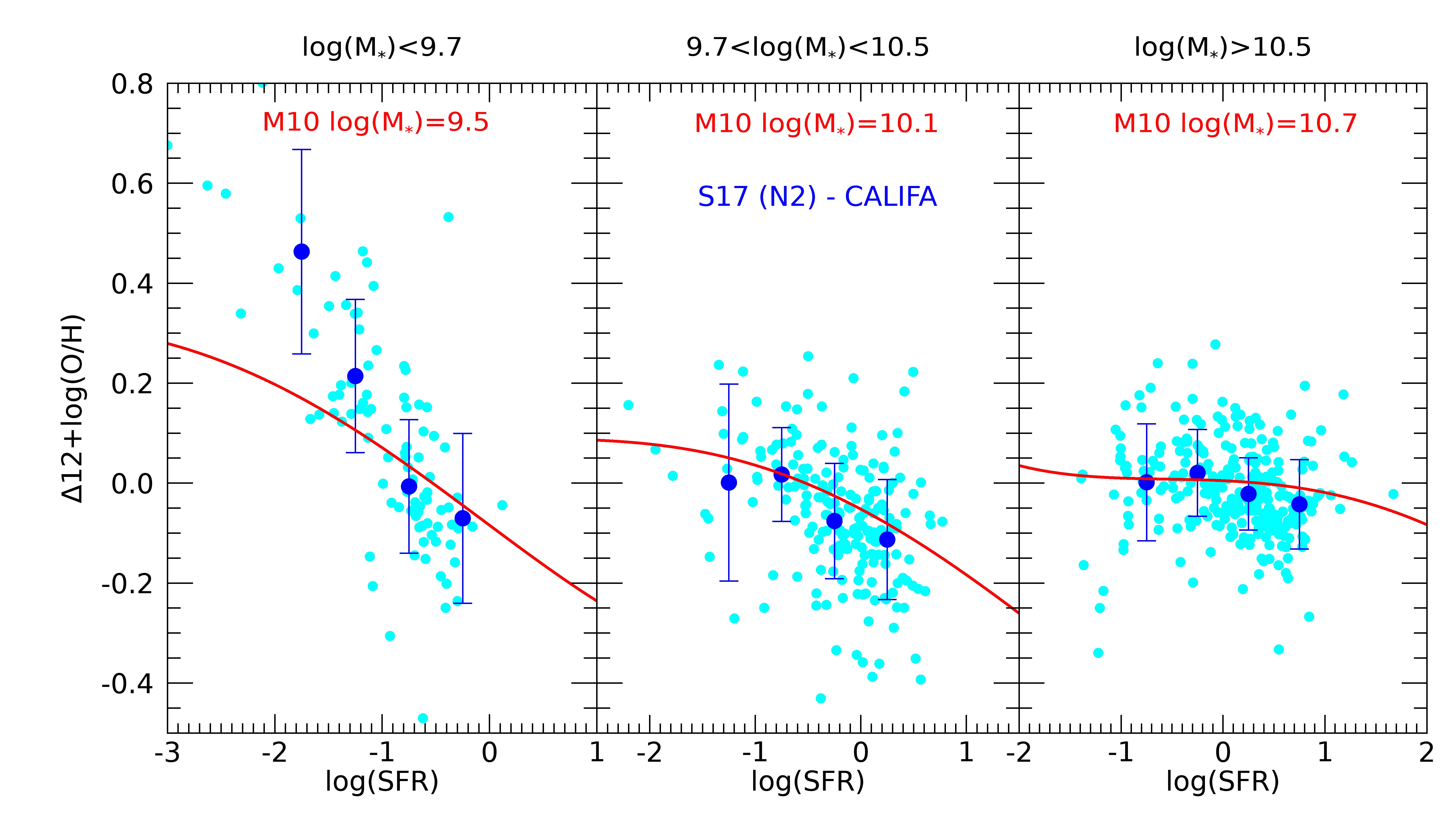}
        \end{center}
        \caption{MZ residuals for the CALIFA galaxies in S17, rescaled to the Maiolino et al. \cite{maiolino08} N2 calibration, plotted against their SFR in three mass bins. The cyan circles show single galaxies, while the blue points indicate the median in SFR bins with their $1\sigma$ dispersion. The red lines indicate the expectations of the FMR for a representative $M_*$ in each bin from M10. Clearly a strong dependency of SFR is present at low mass, decreasing in the higher mass bins, in agreement with the predictions of the FMR.} 
        \label{califamassbin}
\end{figure}

The MZ relation for the final CALIFA sample of 612 galaxies was later presented in S\'anchez et al. (\citealp{sanchez17}, S17). These authors used the same approach as S13, deriving metallicities at the same physical scale $R_{eff}$ for all the galaxies and obtaining a MZ relation with similar shape for the various calibrators used. Again, they claimed that no significant secondary relation of the MZ with either the SFR or the sSFR is present in the residuals of the best-fitting MZ relation, except at low masses $M_*<10^{9.5}\ M_{\odot}$, where some correlation is found but the sample is not complete. However, this is again because S17 drew their conclusion on the analysis of their Fig.~4, where all masses are plotted together, reducing the dependency on SFR at fixed mass below what can be detected given their small sample. 
Actually, the flat relation they derive is in perfect agreement with the SDSS M10 sample. This is in fact shown in Fig.~\ref{allmass}, upper panel, where we plot in red the deviation from the MZ relation (i.e. the difference between the metallicity measured for each galaxy and the expected metallicity given its mass from the median MZ of the sample) of the SDSS sample selected in M10 as a function of both SFR and sSFR, including all masses together. The S17 data using their N2 calibration of Marino et al. \cite{marino13} are overplotted in green on the SDSS distribution. The result does not change if a different calibrator is used for the S17 data because the shape of the relation is very similar. 
It can be seen how the relation for the SDSS M10 sample, where the dependency on SFR at fixed mass is very clear (see e.g. Fig.~\ref{yates}), is fully consistent with CALIFA galaxies once the highest sSFR galaxies with $log(sSFR)>-9.6$ are removed from the SDSS sample to match the CALIFA sSFR distribution (panels a and c). 
The standard errors of the mean cannot be reliably used in this work because they do not take into account the differences in sample selection, measurements of mass, SFR, line fluxes, and metallicities, etcetera. Therefore they represent an underestimation of the real uncertainties; for example, in CALIFA  these uncertainties are as small as $\sim0.012$ dex. Nevertheless, even in this case the two samples would be consistent at $2\sigma$ level.

The presumed difference between the two samples is reinforced in S17 by a wrong expectation of the FMR predictions. In their Fig.~4 a blue dashed curve is supposed to show the relation between the MZ residuals and the SFR expected when using the secondary relation proposed by M10. However, that curve rises at low SFR up to $\Delta (log(O/H))\sim0.14$ above the MZ, and shows no difference at high SFR, contrary to the expectations of M10 discussed above and shown in Fig.~\ref{allmass}. In fact, S17 derived that curve based on several wrong assumptions to remove the different Z-SFR dependency at different masses, such as that all the galaxies in the M10 sample follow the main sequence relation between $M_*$ and SFR with no scatter, the galaxies on the MZ relation all have $log(SFR)=0$ and the $\mu_{0.32}$ fit presented in their eq. 4 by M10 can be used as a proxy of the MZ once the SFR factor in $\mu_{0.32}=log(M_*)-0.32\cdot log(SFR)$ is put to 0, and that the two samples span a comparable range in $M_*$, SFR and sSFR. In reality, the dependency on SFR of the MZ residuals predicted by the FMR changes widely as a function of stellar mass because it is very steep at low masses and almost flat for $log(M_*)>10.5$ (see Fig.~\ref{yates}, upper left panel). Given the different dependency expected for different masses, the strong metallicity variation up to $\sim0.4$ dex as a function of SFR at fixed mass is mixed away and cancelled once all masses are plotted together, except for the highest sSFR, as shown above. \\
\begin{figure*}
        \begin{center} 
                \includegraphics[width=0.9\textwidth]{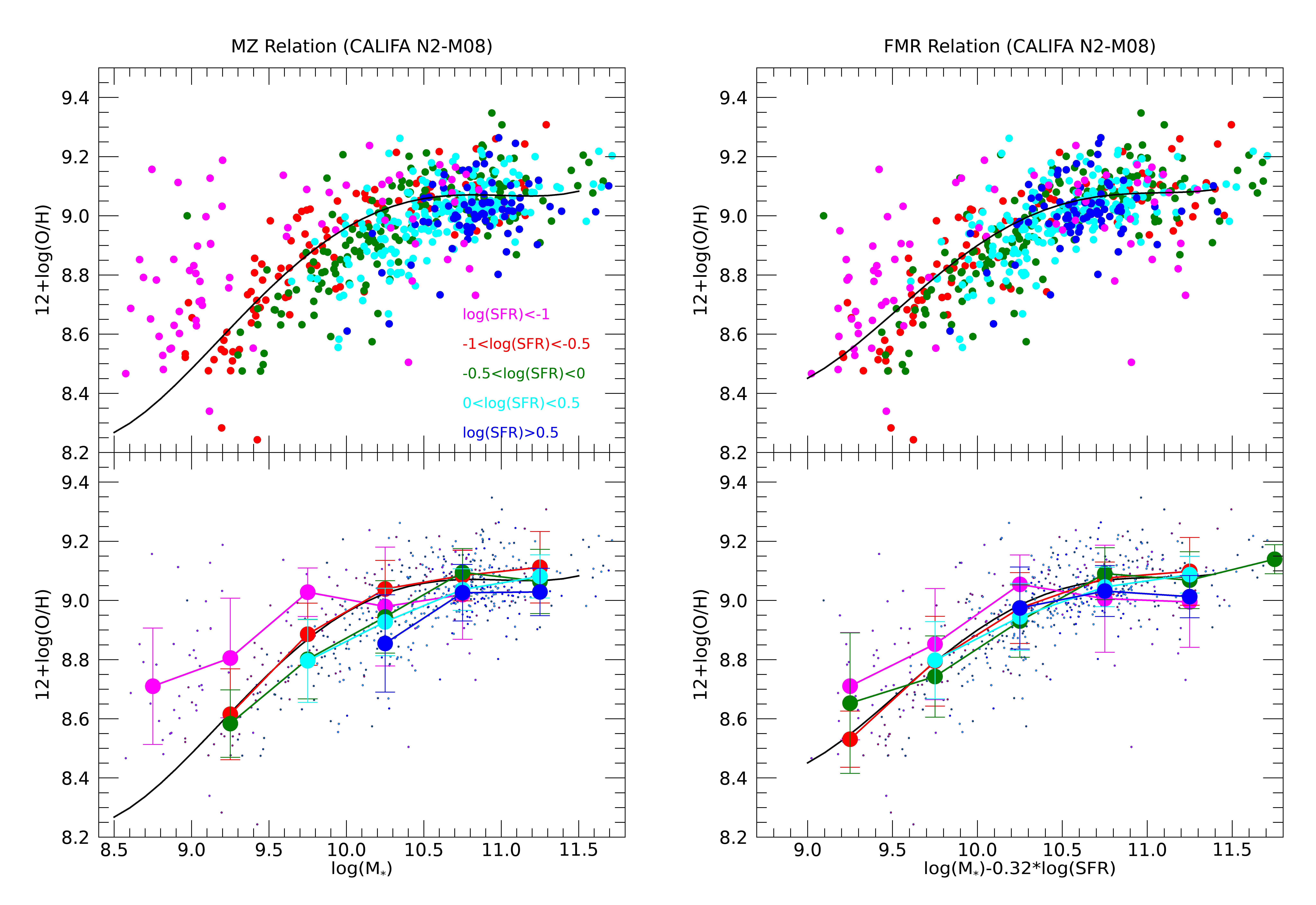}
        \end{center}
        \caption{MZ relation (\textit{left panels}) and FMR projection $\mu_{0.32}$ (\textit{right panels}) of the CALIFA galaxies in S17, rescaled to match the Maiolino et al. \cite{maiolino08} N2 calibration, for a correct comparison with M10 whose MZ and FMR projection for SDSS galaxies are shown as black lines. The CALIFA galaxies are shown with a colour code depending on their SFR, showing that higher SFR galaxies show lower metallicities at fixed mass, as predicted by the FMR. This secondary dependency is removed once the galaxies are plotted on the FMR $\mu_{0.32}$ projection on the left. The lower panels show the medians in each SFR and $M_*$ bins and their $1\sigma$ dispersions, with the same colour code. } 
        \label{califamzfmr}
\end{figure*}

To properly recover the expected FMR dependency on SFR, the galaxies have to be plotted in mass bins, or the MZ relations for the different SFR bins have to be compared. Moreover, consistent metallicity calibrations should be used to allow a proper comparison with the FMR relation of M10. The first approach is shown in Fig.~\ref{califamassbin}, where the CALIFA data from S17 are divided in three mass bins, in which metallicities are rescaled to match the Maiolino et al. \cite{maiolino08} N2 calibration. A clear trend with SFR is present once the galaxies are plotted in this way, especially in the lowest mass bin ($log(M_*/M_{\odot})<9.7$). The trend becomes shallower in the central bin ($9.7< log(M_*/M_{\odot})<10.5$) and almost flat in the highest mass bin ($log(M_*/M_{\odot})>10.5$), which agrees with the FMR expectations (red lines in Fig. 4). \\
The second approach, i.e. plotting the galaxies in SFR bins, was already used by S17 in their Fig.~6, where the MZ distribution for the CALIFA final sample is shown by colour coding the galaxies as a function of their SFR. The best-fit MZ relation for each SFR is overplotted, showing a clear trend of lower metallicities at fixed mass for increasing SFR, exactly as expected by the FMR. Also S17  showed that the dispersion around the MZ in these SFR bins decreases with respect to the global bins, especially at $log(SFR)>0$, where it is reduced to $\sigma\sim0.04$ from the original quoted $\sigma=0.064$ using their \textrm{t2} calibration. Surprisingly, S17 concluded that despite these findings they cannot confirm the existence of a significant secondary dependency of the MZ with the SFR, or at least that it is confined only to $M_*<10^{9.5}\ M_{\odot}$.
We tried the same approach in Fig.~\ref{califamzfmr}, where the CALIFA data from S17 are rescaled to the Maiolino et al. \cite{maiolino08} N2 calibration to compare with the original M10 MZ and FMR relations from SDSS, adding a shift of 0.1 dex to account for the different aperture used to derive the metallicity. We confirmed S17 in finding a clear trend for lower metallicity at higher SFR at fixed mass, that is removed once the galaxies are plotted on the $\mu_{0.32}$ FMR projection, in agreement with the FMR expectations. The scatter of the medians in the different bins significantly decreases from 0.11 dex in the MZ relation to 0.06 in the FMR projection.  \\

We therefore conclude that, consistent with the early work by Salim et al. \cite{salim14} and Fig. 6 of S17, the CALIFA galaxies show a convincing dependency of the MZ residuals against SFR, contrary to the claim of S13 and S17.

\section{Fundamental metallicity relation in MaNGA galaxies} \label{barrera}

The ongoing SDSS-IV MaNGA survey (Bundy et al. \citealp{bundy15}) will investigate the internal kinematical structure and composition of gas and stars in an unprecedented sample of 10000 nearby galaxies, using fibre-fed integral field spectroscopy. Barrera-Ballesteros et al. (\citealp{barrera17}, hereafter BB17) used the first targets observed until June 2016 (2730 galaxies at $0.03<z<0.17$) to study their MZ relation. Following the S13 approach, they used for each star formation dominated galaxy the metallicity at $R_{eff}$, as derived interpolating a radial metallicity gradient of selected HII regions-like spaxels and applying ten different calibrators. The obtained  MZ relation shape is consistent with previous works in local galaxies, while the authors claim that no secondary relation is present between the relation residuals and SFR, contrary to the FMR expectations.\\
To demonstrate their claim, BB17 plot in their Fig. 3 the metallicity residuals of the MZ relation against the SFR of all the galaxies together, finding a basically flat curve for all the calibrators adopted. As for the CALIFA sample, this was compared with a presumed M10 expectation, which clearly does not reproduce the data. As discussed above, when galaxies are not divided according to mass, the metallicity dependency on SFR is $\lesssim\pm0.05dex$ even in the M10 sample, and the three samples (M10, S17 and BB17) follow the same relation. The true expectation from the SDSS M10 sample based on the FMR is plotted in Fig.~\ref{allmass}, where the BB17 data points for the N2 calibration of Marino et al. \cite{marino13} are overplotted in blue. Again, once the highest $log(sSFR)>-9.6$ galaxies in the M10 sample are removed to match with the MaNGA sample,  agreement is reached between the two datasets. The result does not change if a different calibrator is used for the BB17 data, as the shape of the relation is very similar.  Even assuming standard errors of the mean instead of standard deviations, the two distributions are always consistent within 2$\sigma$ or 0.01 dex.\\

\begin{figure}
        \begin{center}
                \includegraphics[width=0.5\textwidth]{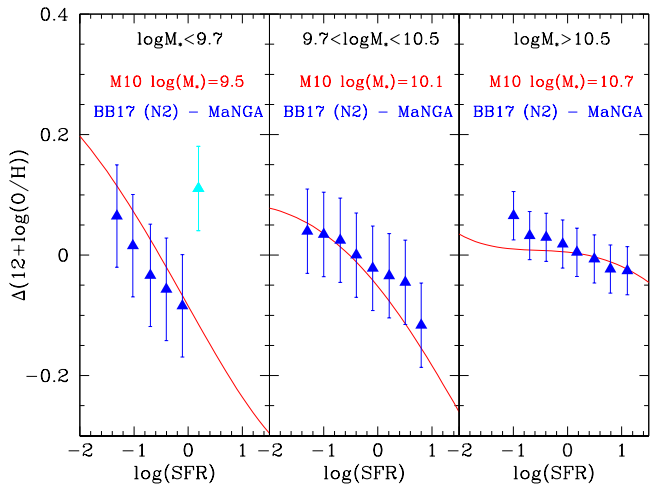}
        \end{center}
        \caption{MZ residuals for the MaNGA galaxies in BB17 using the N2 calibration against their SFR in three mass bins, shown as blue triangles, derived from Fig.~5 of BB17. The cyan point indicates the highest SFR bin at low mass, where the fraction of galaxies in BB17 is low and the scatter among different calibrators is much higher (see discussion in BB17). The red lines show the expectations of the FMR, directly for eq.~4 of M10. The error bars represent the dispersion of the residuals in each bin (see BB17). Contrary to what was affirmed by BB17, MaNGA galaxies closely follow the prediction of the FMR in M10 in all mass bins.} 
        \label{mangabins}
\end{figure}

Interestingly, BB17 also show the MZ residuals as a function of SFR divided in three mass bins in their Fig. 5. As expected, in this case a clear trend with SFR is present in the lowest mass bin ($log(M_*/M_{\odot})<9.7$), which becomes shallower in the central bin ($9.7< log(M_*/M_{\odot})<10.5$) and almost flat in the highest mass bin ($log(M_*/M_{\odot})>10.5$); this agrees with the FMR expectations. As shown in Fig.~\ref{mangabins}, 
the expected shape based on the FMR is very well reproduced by MaNGA data once the prediction of M10 from their eq.~4 is plotted for $log(M_*/M_{\odot})=9.5$, $log(M_*/M_{\odot})=10.1$, and $log(M_*/M_{\odot})=10.7$ for the three BB17 bins. We again adopted the N2 calibrator, although a similar result would have been obtained with the others given the similar shape, and we plot in cyan the highest SFR bin in BB17 for $log(M_*/M_{\odot})<9.7$, which shows a large scatter in the different calibrators given the low number of galaxies with measured metallicity (see discussion in BB17).
The BB17 work instead compared the observed trends with the wrong FMR curve presented in S17, as discussed in the previous section. Based on this wrong comparison, BB17 concluded that the MZ relation does not correlate with SFR in MaNGA galaxies, but given the analysis presented above we instead find that the MZ relation scatter of the MaNGA sample studied by BB17 shows a clear dependency on SFR at all masses, consistent with the expectations of the FMR as presented in M10.

\section{Fundamental metallicity relation at high-z} \label{highz}

The M10 work has already highlighted that high redshift galaxies, at least up to $z\sim2.5$, follow the same FMR defined by local SDSS galaxies with no indication of evolution. In this framework, the observed evolution of the MZ relation up to z=2.5 would result from galaxies showing progressively higher SFR; therefore lower metallicities are selected at increasing redshifts, sampling different parts of the same FMR. Although some evolution may be present at higher redshift (Troncoso et al. \citealp{troncoso14}), as expected by some theoretical models (see e.g.  Dav\'e et al. \cite{dave11}), this finding seems to indicate that the same physical mechanisms regulating metallicity mass assembly and SFR are driving galaxy evolution in the last 80\% of cosmic time (Lilly et al. \cite{lilly13}).\\
Distant galaxies show larger dispersions than local SDSS galaxies, between 0.2 and 0.3 dex. At least part of these relatively larger dispersion is due to the larger uncertainties in the estimates of metallicity, mass, and SFR, but part of it could be intrinsic, related to different evolutionary stages of the galaxies. Nonetheless, most high-z galaxy samples with measured metallicities agree with the functional form of the M10 FMR, extrapolated to high SFR (see e.g. Richard et al. \citealp{richard11}, Cresci et al. \citealp{cresci12},  Belli et al. \citealp{belli13}, Maier et al. \citealp{maier14}, Kacprzak, et al. \citealp{kacprzak16}, Sanders et al. \citealp{sanders18}). However, there were claims that some high redshift galaxy samples do not support the presence of a non-evolving FMR. In this section we briefly discuss such claims, showing that large galaxy samples with a wide range in SFRs are needed to directly show a clear dependency on SFR of the MZ scatter at high redshift, given the larger intrinsic spread and uncertainties in high-z measurements.\\

Steidel et al. \cite{steidel14} made use of deep near-IR observations of 251 galaxies at $2.0<z<2.6$ ($2<SFR<200\ M_{*}\ yr^{-1}$, $8.6 < log(M_{*}/M_{\odot}) < 11.4$) observed with the Multi-Object Spectrometer for Infra-Red Exploration
 (MOSFIRE) spectrometer on the Keck 1 telescope in the framework of the Keck Baryonic Structure Survey (KBSS), covering both H and K band, to provide their MZ relation using a recalibrated version of N2 and O3N2 indices. Their best-fit $z\sim2.3$ MZ is somewhat shallower than other studies have suggested, and the intrinsic scatter they derive for the relation ($\sigma_{MZ}\sim0.1$) is similar to what was measured in the local Universe. These authors also investigated whether there was evidence in their sample for a dependency at fixed stellar mass between metallicity and SFR, finding nearly identical best-fit MZ relations in normalization, slope, and intrinsic scatter for two independent subsamples (121 galaxies each), although with median SFRs different by only a factor of $\sim3.4$ (their Fig.~25). Steidel et al. \cite{steidel14} therefore concluded that, at least over the range spanned by their sample, metallicities appear to be independent of SFR at a given stellar mass at $z\sim2.3$. \\
 However, given the small difference in SFR between the two samples, the expected metallicity difference is small, i.e. a median offset of $\sim 0.1$ dex in metallicity. Such a difference is difficult to observe given the measurement uncertainties in all the parameters involved ($M_*$,  O/H and SFR), especially at high redshift, and the possible larger intrinsic scatter compared to the local Universe. Actually, the small median observed difference between the two samples of $\sim0.09$ dex, as derived by our analysis, is perfectly compatible with what is expected based on the FMR. To clearly and directly disentangle the SFR contribution to the scatter in the MZ relation, larger samples of galaxies with a greater dynamical range in SFR are therefore required (see e.g. Sanders et al. \citealp{sanders18}). \\
\begin{figure}
        \begin{center}
                \includegraphics[width=0.5\textwidth]{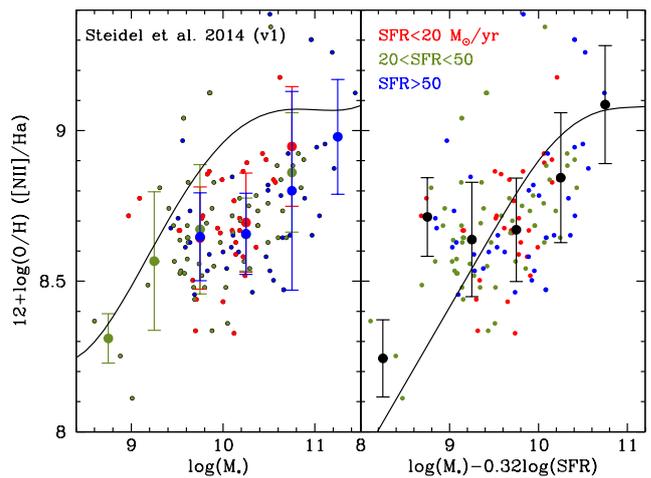}
        \end{center}
        \caption{MZ and FMR relation for the Steidel et al. \cite{steidel14} sample. In the left panel the MZ relation is shown; galaxies are colour coded into three bins based on their measured SFR. The black solid line shows the z=0 M10 MZ relation using SDSS galaxies; the observed offset from the $z\sim2$ galaxies is due to the evolution of the MZ relation. The error bars represent the $1\sigma$ dispersion in each bin. 
	The right panel shows the FMR $\mu_{0.32}$ projection for the 
	same data points; the black points indicate medians in bins of $\mu_{0.32}$, showing that the Steidel et al. \cite{steidel14} $z\sim2.3$ sample is reasonably reproduced by the FMR defined with SDSS galaxies at z=0 (shown in black).} 
        \label{figsteidel}
\end{figure}
\begin{figure*}
        \begin{center}
                \includegraphics[angle=0,width=0.7\textwidth]{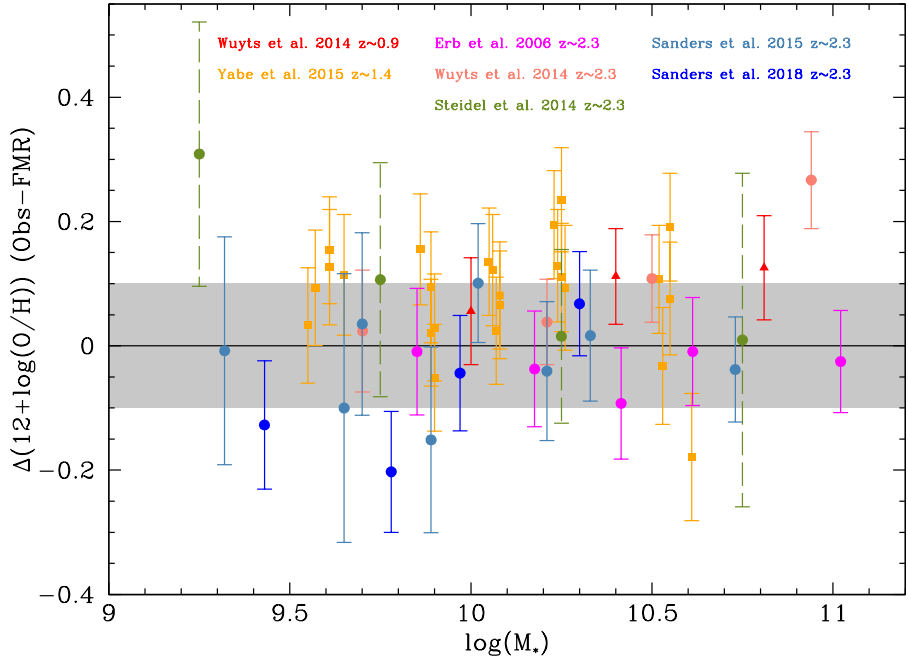}
        \end{center}
        \caption{Offset between the observed metallicity in the different samples of $z\sim0.9$ and $z\sim2.3$ galaxies and the predicted metallicity based on their $M_*$ and SFR, according to the FMR defined by SDSS galaxies at $z\sim0$ by eq. 1 in M10 and Mannucci et al. \cite{mannucci11}. All the data points are based on single measurements on stacked spectra in mass and SFR bins; the error bars show the corresponding metallicity uncertainty, except for Steidel et al. \cite{steidel14}, whose medians of single galaxies in mass bins are plotted with their dispersion as a dashed line. The grey area shows the $\pm0.1$ dex region around 0, which represents an estimate of the uncertainty tied to the FMR extrapolation and parameter estimation in the range covered by the data points (see Curti et al. in prep). Once the FMR and the measured metallicities are plotted on a consistent scale (in this case Maiolino et al. \cite{maiolino08} calibration), no apparent evolution of the FMR up to $z\sim2.3$ is found.} 
        \label{fmrdiff}
\end{figure*}

A much better test of the validity of the FMR in high-z samples is instead the comparison of the position of these $z\sim2.3$ galaxies on the FMR surface defined at $z=0$, or on its $\mu_{0.32}$ projection (see e.g. Cresci et al. \citealp{cresci12}), thanks to a much larger SFR difference between SDSS and high-z galaxies. In this way it can be verified if the FMR defined at z=0.1 by the SDSS is capable of predicting the average metallicity of high-z samples. We therefore performed this test using the subsample of 130 galaxies presented in the first version of the Steidel et al. \cite{steidel14} paper submitted on astro-ph\footnote{uploaded on May 21, 2014, available at https://arxiv.org/abs/1405.5473v1}, where emission line ratios, stellar masses, and SFRs are tabulated, as the SFR measurements are not included in the final version of the paper for the full sample. In Fig.~\ref{figsteidel} the MZ relation for this subsample is shown using the Maiolino et al \cite{maiolino08} calibration (left panel); galaxies are divided into three bins based on their SFR. A small difference in metallicity at fixed mass is visible between these three bins, although with low significance given the large intrinsic scatter and the small dynamical range in SFR. In the right panel the projection $\mu_{0.32}$ of the FMR for medians in $\mu_{0.32}$ of the subsample is plotted, showing good agreement with the $z\sim0$ relation defined by SDSS galaxies in M10. We therefore conclude that the Steidel et al. \cite{steidel14} $z\sim2.3$ sample also shows metallicities consistent with the expectations of the FMR given their SFR, and that a comparison with the $z\sim0$ FMR has to be performed to verify the applicability of the relation at high-z, especially in small samples.\\

Wuyts et al. \cite{wuyts14} presented the correlations between stellar mass, SFR, and gas-phase metallicity for a sample of 222 galaxies at $0.8 < z < 2.6$, observed with Luci at the Large Binocular Telescope (LBT) as well as Spectrograph for INtegral Field Observations in the Near Infrared (SINFONI) and K-band Multi Object Spectrograph (KMOS) at the Very Large Telescope (VLT). These authors used the  [NII]/H$\alpha$ flux ratio as a proxy for metallicity, finding larger redshift evolution at lower masses. Interestingly, their $z\sim2$ MZ relation is much steeper than that derived by Steidel et al. \cite{steidel14} and discussed above. In fact, the Wuyts et al. relation in the lowest mass bin, $log(M_*/M_{\odot})\sim9.6$, is $0.18$ dex lower in metallicity than the Steidel et al. \cite{steidel14} relation (see their Fig. 2, right panel). This offset can be fully explained in the FMR framework, as in that mass bin the two samples have a different median SFR, i.e. $SFR\sim40\ M_{\odot}/yr$ for Wuyts et al. \cite{wuyts14} and  $SFR\sim16\ M_{\odot}/yr$ for Steidel et al. \cite{steidel14}. At that low mass, where the FMR effect is higher for such a SFR offset, the FMR predicts exactly the observed 0.18 dex difference between the two MZ relations. This clearly shows how any observed MZ relation strongly depends on the sample used and on its SFR distribution.\\ 
To test their data against the FMR, Wuyts et al. divided the samples at $z=0.9$ and $z=2.3$ into two SFR bins at each stellar mass, finding at both redshifts no clear difference between the [NII]/H$\alpha$ ratio at fixed mass between the high and low SFR bin. However, we calculated that the small expected difference between the two samples based on the FMR (i.e. median of $\sim 0.1$ dex in metallicity at both $z=2.3$ and $z\sim0.9$ with a maximum of $0.16$ dex) is easily washed away by the measurements uncertainties and intrinsic scatter, as just the uncertainty on metallicity is larger or comparable, and therefore difficult to detect. 
As for the Steidel et al \cite{steidel14} sample, the median metallicity difference between the MZ derived locally and the MZ derived at $z\sim 0.9$ and $z\sim2.3$ is much larger than the intrinsic scatter at fixed redshift ($\sim0.6$ dex in metallicity and $\sim 0.3$ dex for $z\sim 2.3$ and $z\sim0.9$ for the lowest mass bin, respectively), and therefore a more meaningful test of the FMR at high-z should be the position of the high-z galaxies on the FMR surface defined at $z=0$, or on its $\mu_{0.32}$ projection. This test is shown in their Fig.~4, where they in fact find an overall agreement between the high-z samples and the local FMR in M10.\\

Sanders et al. \cite{sanders15} presented the first results based on a first sample of 87 $z\sim2.3$ galaxies from the MOSFIRE Deep Evolution Field (MOSDEF) survey, using the multi-object near-IR spectrograph MOSFIRE at Keck 1 telescope, finding a well-defined MZ relation with an offset of $\sim0.15-0.3$ dex below the local one defined by SDSS. Following Steidel et al. \cite{steidel14} and Wuyts et al. \cite{wuyts14}, they also divided the sample into two SFR bins, concluding that no significant difference in the MZ relations of the two subsamples was present. A possible metallicity segregation between the two samples is already visible in their Fig.~4, in which more active galaxies generally show lower metallicities at fixed mass. However, from the tabulated values of Sanders et al. \cite{sanders15} a mean difference of $\sim0.15$ dex is actually present between the high and low SFR bins, but with the higher SFR bins showing higher metallicity; this is naively opposite from the expected FMR trend. This offset has to be compared with the expected difference of $\sim0.1$ dex in the same direction based on the FMR predictions using the Maiolino et al. \cite{maiolino08} calibrations. In fact, the metallicity trend expected is fully dominated by the mass difference between the high and low SFR bins ($\sim0.5$ dex in $M_*$), and not by the different SFR in the two bins: while the different SFR produces an effect of less than 0.1 dex with higher SFR galaxies showing lower metallicities, the mass difference dominates and inverts the trend, in which more massive galaxies show higher metallicities. This shows again how the effect introduced by SFR has to be carefully evaluated, especially at high-z. Sanders et al. concluded that there may still be SFR dependency of the MZ relation at $z \sim 2.3$, although their current sample lacks the size and dynamic range required to resolve it. In fact, once the larger sample of 260 galaxies from MOSDEF is analysed in Sanders et al. \cite{sanders18} by the same group, they find that a $M_*$-SFR-Z relation clearly exists at $z\sim2.3$, although they find metallicities lower by $\sim0.1$ dex at fixed $M_*$ and SFR when compared to local analogues by Andrews \& Martini \cite{andrews13}. \\

However, this offset disappears once the metallicities are compared with the FMR in M10 using the same Maiolino et al. \cite{maiolino08} metallicity scale. This is shown in Fig.~\ref{fmrdiff}, where we plot the offsets between the FMR predictions and the measured metallicities in Sanders et al. \cite{sanders18}, the other high-z samples discussed previously, and other $z>1$ samples in the literature. The points from Erb et al. \citealp{erb06} are based on stacked spectra in mass bins from a sample of 87 Lyman break galaxies at $z\sim2.3$ observed with NIRSPEC at Keck II, while the Yabe et al. \cite{yabe15} data are stacked spectra in mass and SFR bins from a large survey of $\sim1400$ galaxies at $z\sim1.4$ with Fiber Multi-Object Spectrograph (FMOS) at Subaru. For this Yabe et al. \cite{yabe15} sample, a fixed aperture correction to the H$\alpha$ fluxes used to compute the SFRs was removed, for consistency with M10 who did not apply aperture corrections. The shift by $\sim0.15$ dex from the local FMR found by Yabe et al. \cite{yabe15} with this dataset was attributed to the increase in the nitrogen-to-oxygen abundance ratio at fixed metallicity at $z\sim1.4$, which may bias the metallicity determination based solely on the [NII]/H$\alpha$ indicator. Although this effect might be present, the offset was probably partly due to the uncertainties of this fixed aperture correction, as the difference is significantly decreased to $\sim0.08$ dex once the correction is removed. We also note that the significant uncertainties tied to the large extrapolation of the FMR covered by SDSS galaxies up to the high SFR of these high-z samples may also contribute towards increasing the scatter and shifting the overall normalization, up to $\sim 0.2$ dex at $log(SFR)>2.5$ (see Curti et al. in prep). The typical extrapolation uncertainty for the mass and SFR range considered in this work ($\sim0.1$ dex), is plotted as a grey area, derived through a Monte Carlo Markov Chain by letting the different FMR parameters derived according to the Curti et al. (in prep.) parameterization to randomly vary within their uncertainties. This value also corresponds to the average spread of the metallicity calibrations. Therefore, the residual offset in the Yabe et al. \cite{yabe15} dataset is probably connected to the large extrapolation uncertainty, as this sample is characterized by the highest SFR among those considered in this work.\\

Once all the high-z samples are considered together and both the FMR parametrization and metallicities are converted to the same scale, in this case Maiolino et al. \cite{maiolino08}, no significant evolution
of the relation is present. The median offset of all the data points in Fig.~\ref{fmrdiff} with respect to the local FMR is 0.06 dex, where $\sigma=0.1$, thus consistent with no evolution. For comparison, their median offset from the local MZ is -0.29 dex.  
The remarkable agreement of all these high-z datasets with the FMR defined by local galaxies confirms that this relation shows no evolution up to $z\sim2.3$, at least in the samples analysed.

\section{Conclusions} \label{conclusion}

We have compared the different functional forms of the three-parameter relation between stellar mass $M_*$, SFR, and gas-phase metallicity Z in local star-forming galaxies to understand the origin of the differences presented in several works. Despite the different shapes obtained, mostly as a consequence of the adopted metallicity calibration and SFR indicator, all the authors of the works mentioned have confirmed that these three properties are closely connected and that the scatter of the MZ relation can be reduced by using the SFR as third parameter ($\sigma=0.05$ for SDSS-DR7 galaxies). \\
We also reanalysed the data of CALIFA and SDSS-IV MaNGA samples, which were used in the literature by S13, S17, and BB17 to claim that no secondary relation was present between the MZ relation and SFR. We find that those datasets are consistent with the FMR scenario as well, as  in these samples the metallicity at fixed mass also depends on the SFR.\\
Although several works suggest that the FMR also holds  for high redshift galaxies, at least up to $z\sim2.5$, some samples were reported to be inconsistent with its existence at high-z (Steidel et al. \citealp{steidel14}, Wuyts et al. \citealp{wuyts14}, Sanders et al. \citealp{sanders15}). We reanalysed these data, finding that contrary to previous claims the data  are  in good agreement with the $z\sim0$ FMR, showing that the relation still holds up to $z\sim2.5$. We note that large samples with a high dynamical range in SFR are required to internally detect the effect of SFR in high-z MZ relations, and thus a more meaningful test of the FMR at high-z is usually compared with the local FMR defined by SDSS galaxies. 
We also stress the importance of comparing metallicity measurements and FMR predictions on the same metallicity scale to avoid the introduction of offsets due to the different calibrations adopted.\\
The confirmed existence of this 3D relation up to high-z has an important implication for our picture of galaxy evolution, suggesting a long-lasting equilibrium between gas inflows, outflows, and star formation in galaxies (see e.g. M10, Lilly et al. \citealp{lilly13}), and it thus represents a fundamental test bench for any galaxy evolution model.

\section*{Acknowledgments}

GC is grateful to Sebastian S\'anchez and Jorge Barrera-Ballesteros for their useful discussion. GC acknowledges the support by INAF/Frontiera through the "Progetti Premiali" funding scheme of the Italian Ministry of Education, University, and Research; GC and FM have been also supported by the INAF PRIN-SKA 2017 programme 1.05.01.88.04. MC acknowledges support from the ERC Advanced Grant 695671 "QUENCH” and support from the Science and Technology Facilities Council (STFC). 
This study uses data provided by the Calar Alto Legacy Integral Field Area (CALIFA) survey (http://califa.caha.es/). Based on observations collected at the Centro Astronómico Hispano Alemán (CAHA) at Calar Alto, operated jointly by the Max-Planck-Institut f\"ur Astronomie and the Instituto de Astrofísica de Andalucía (CSIC). Funding for the Sloan Digital Sky Survey has been provided by the Alfred P. Sloan Foundation, the U.S. Department of Energy Office of Science, and the Participating Institutions. GC and FM finally acknowledge D. Lucano's, P. Bartolo's and M. Biancalani's standing for laws that "not today or yesterday but always they live".


\end{document}